# Quantum-like behavior of 1D nonequilibrium system in the maximum heat flux limit

S. L. Sobolev[a,b]

[a] Institute of Problems of Chemical Physics, Academy of Sciences of Russia, Chernogolovka, Moscow Region, 142432 Russia

[b] Samara State Technical University, ul. Molodogvardeiskaya 244, Samara, 443100 Russia

E-mail: sobolev@icp.ac.ru

## Abstract

Using information entropy formalism, we consider a one-dimensional system with heat flux and extend the meaning of equilibrium variablesto nonequilibrium scenarios when classical local equilibrium approach is not applicable; this is particularly important for the performance evaluation of modern thermal systems and microdevices, which usually operate in extreme situations. The extended nonequilibrium entropy, temperature, thermal conductivity and heat capacity have been analyzed as functions of the local energy density (kinetic temperature) and heat flux. The ratio of the heat flux to its maximum possible value plays a role of an order parameter – it varies from zero in the equilibrium, i.e. thermalized (disordered) state, to unity in the nonequilibrium (completely ordered) state. We demonstrate that there are some analogies between the behavior of the nonequilibrium systems in the maximum heat flux limit and equilibrium quantum system in the low temperature limit, whicharise due to the breakdown of the equipartition in both cases. This implies that there can be fruitful cross-fertilization of ideas and techniques between these two fields.

## I. INTRODUCTION

Classical thermodynamics has been highly successful, impacting strongly on the natural sciences and enabling the development of technologies that have changed our lives, from fridges to jet planes. Until recently, it was applied to large local equilibrium systems described by the laws of classical physics. However, with modern technologies miniaturizing down to the nanoscale and into the quantum regime, testing the applicability of thermodynamics in this new realm has become an exciting technological challenge. As a result the fields of non-equilibrium thermodynamics and quantum thermodynamics have recently started to blossom, fuelled by new, highly controlled experiments and the availability of powerful numerical methods, such as molecular dynamic simulations.

A challenge in nonequilibrium thermodynamics is extending the meaning of equilibrium variables to nonequilibrium scenarios. Classical irreversible thermodynamics (CIT) deals with near-equilibrium situations and assumes that the local thermodynamic equilibrium (LTE) conditions prevail. In this situation the concepts of equilibrium thermodynamics, including variables such as temperature and entropy, can be defined as in global equilibrium and are applicable locally. However, our interest is in the transport of heat through nanoscale systems and/or on ultra-short time scale. Can temperature, which is an equilibrium concept, still be invoked in a nonequilibrium process such as heat flow? The question what precisely is a "local temperature" under far from LTE conditions is a subject of intense theoretical and experimental interest in a broad context of physics, chemistry and life sciences and is open to discussion [1-16].

Molecular dynamic (MD) simulations, which are often used to study heat flow under far from equilibrium conditions, usually define temperature on the bases of an average kinetic energy [1,2,10,14,23-38]. However, at low temperatures, when the quantum effects come into play, the definition needs quantum correction [23-37]. Maximum entropy formalism (MEF) [9], extended irreversible thermodynamics (EIT) [3], thermomass model (TMM) [5], and information entropy formalism (IEF) [6,7] define the nonequilibrium temperature $\theta$ by analogy with the classical definition $\theta = (\partial S_{neq} / \partial E)^{-1}$, where $S_{neq}$ is the local nonequilibrium entropy, $E$ is the local energy density. Some analogies between the behavior of equilibrium systems at low absolute temperature and nonequilibrium system under high values of the heat flux have been observed [9]. For glassy systems, the definition of the effective temperature is based on a modified version of the fluctuation–dissipation theorem (FDT) [3,7]. In the active systems the generalized effective temperature is a product of two factors: the first one arises due to the system activity and describes the randomized motion of the particles, whereas the second one is a consequence of the ordered (collective) motion due to the alignment effects or initially ordered configuration [7]. An additional micromorphic temperature associated with the classical local temperature is introduced as an independent degree of freedom, based on the generalized principle of virtual power [13]. An effective temperature has also been introduced in disordered semiconductors under electric field when one can characterize the combined effects of the field and the lattice temperature by an effective temperature to describe carrier drift mobility, dark conductivity, and photoconductivity [7]. Despite these advances, the question how to treat temperature under far from LTE conditions is open to discussion.The problem is that in practical situations, the routine analysis of nonequilibrium heat transport phenomena has been limited by the high computational demand of rigorous simulations.

Quantum thermodynamics is an emergent research field, which combines ideas from non-equilibrium thermodynamics, open quantum system theory, quantum information and mesoscopic physics. While many different approaches have been put forward in the last years to understand thermodynamics in the non-equilibrium and quantum regimes, a remaining fundamental question to be answered is to what extent the paradigms of local equilibrium thermodynamics can be used by the non-equilibrium and quantum thermodynamics. Moreover, it is important to bridge the gap between the non-equilibrium and quantum approaches, which goals have much in common, ranging from the desire to give new interpretation of the laws of thermodynamics to the need of designing useful non-equilibrium quantum heat engines in a new era of increased nanotechnological abilities.

The main purpose of this paper is to demonstrate and discuss the analogies in the behavior of nonequilibrium systems under high values of the heat fluxand equilibrium quantum system at low temperatures. In section 2 we analyze the extended nonequilibrium variables, such as the entropy, temperature, heat capacity, and thermal conductivity, as functions of the local energy density and the heat flux, which characterizes the extent of deviation from equilibrium.In section 3 we compare the behavior of the extended nonequilibrium variables in the maximum heat flux limit with the behavior of their equilibrium counterparts in the low temperature quantum limit. Conclusions are made in section 4.

## 2. Model and results

*2.1 Governing equations*

The Boltzmann transport equation (BTE) with the single relaxation time (or BGK) approximation is given by [2,4,7,36-38]

$$\frac{\partial f}{\partial t} + \vec{v} \cdot \vec{\nabla} f = -\frac{f - f^0}{\tau} \quad (1)$$

where $f$ is the phonon distribution function, $\vec{v}$ is the phonon group velocity, and $f^0$ is the equilibrium distribution function. BTE, Eq.(1), can be cast into an equation for the phonon energy density $E$ by integrating it over the frequency spectrum. For simplicity, the effects of temperature on the dispersion relations and the phonon density of states are neglected. Then, the BTE, Eq.(1), in a phonon energy density formulation for 1D is given by [7,21]

$$\frac{\partial E}{\partial t} + v_x \frac{\partial E}{\partial x} = -\frac{E - E^0}{\tau} \quad (2)$$

where $E^0$ is the equilibrium phonon energy density, and $v_x$ is the component of velocity along the $x$-axis. Since in 1D the phonons can travel in the positive or negative direction along the $x$-axis, Eq.(2) gives two equations

$$\frac{\partial E_1}{\partial t} + v \frac{\partial E_1}{\partial x} = -\frac{E_1 - E_1^0}{\tau} \quad (3)$$

$$\frac{\partial E_2}{\partial t} - v \frac{\partial E_2}{\partial x} = -\frac{E_2 - E_2^0}{\tau} \quad (4)$$

where $E_1(x,t)$ is the energy density of carriers moving to the right, $E_2(x,t)$ is the energy density of carriers moving to the left, $E_i^0$ is the corresponding equilibrium energy densities. The sum and the difference of Eqs.(3) and (4) give, respectively

$$\frac{\partial E}{\partial t} = -\frac{\partial J}{\partial x} \quad (5)$$

$$J + \tau \frac{\partial J}{\partial t} = -D \frac{\partial E}{\partial x} \quad (6)$$

where $J = v(E_1 - E_2)$ is the energy flux, $E = E_1 + E_2$ is the energy density, $D = v^2 \tau$ is the thermal diffusivity. Eq.(5) is the energy conservation law, while Eq.(6) is the analog to the modified Fourier law with allowance for the relaxation to local equilibrium with characteristic relaxation time $\tau$.

Equations for $E_i$ in terms of the energy density $E$ and the energy flux $J$ take the form

$$E_1 = (E + J/v)/2 \quad (7)$$

$$E_2 = (E - J/v)/2 \quad (8)$$

Note that the governing equations for $E_i$ are similar to Eq.(7), i.e. are of hyperbolic type.

*2.2 Information entropy and nonequilibrium temperature*

The information entropy is given by [38,39]

$$S_{neq} = -\sum_i u_i \ln u_i \quad (9)$$

where $u_i$ is the distribution function of subsystem $i$. For the system under consideration we have two subsystems ($i$=1,2), which distribution functions can be represented in terms of the corresponding energy densities as $u_i = E_i/E$. Eq.(9) allows us to represent the nonequilibrium entropy $S_{eq}$ in terms of energy densities as follows [7]

$$S_{neq} = -[E_1 \ln(E_1/E) + E_2 \ln(E_2/E)]/E \tag{10}$$

Using Eqs.(7) and (8) for $E_i$, the nonequilibrium entropy, Eq.(10), can be written in terms of energy flux $J$ and energy density $E$ as

$$S_{neq} = -\frac{1}{2}(1+J/vE)\ln\frac{1}{2}(1+J/vE) - \frac{1}{2}(1-J/vE)\ln\frac{1}{2}(1-J/vE) \tag{11}$$

After some algebra, Eq.(11) can be represented as

$$S_{neq} = S_{eq} - \frac{1}{2}(1+J/vE)\ln(1+J/vE) - \frac{1}{2}(1-J/vE)\ln(1-J/vE) \tag{12}$$

where $S_{eq} = \ln 2$ is the equilibrium entropy. The Lagrange multipliers, which correspond to the energy and the heat flux constraints, are given as [9]

$$\beta = \left(\frac{\partial S_{neq}}{\partial E}\right)_J \tag{13}$$

$$\gamma = \left(\frac{\partial S_{neq}}{\partial J}\right)_E \tag{14}$$

After some calculations, we obtain from Eqs.(12)-(14) expressions for $\beta$ and $\gamma$ in terms of the macroscopic variables $E$ and $J$ as follows

$$\beta = \frac{\partial S_{eq}}{\partial E} + \frac{J}{2vE^2}\ln\frac{1+J/vE}{1-J/vE} \tag{15}$$

$$\gamma = \left(\frac{\partial S_{neq}}{\partial J}\right)_E = \frac{1}{2vE}\ln\frac{1-J/vE}{1+J/vE} \tag{16}$$

By analogy with equilibrium thermodynamics, the generalized (nonequilibrium) temperature $\theta$, which depends not only on the energy density $E$ but also on the heat flux $J$, is defined as $\theta = 1/\beta$ [7,9]. Thus, Eq.(15) gives the expression for the nonequilibrium temperature $\theta$, which generalizes the classical thermodynamic definition $T = (\partial S_{eq}/\partial E)^{-1}$ to the nonequilibrium case:

$$\frac{1}{\theta} = \frac{1}{T_k} + \frac{J}{2vcT_k^2}\ln\frac{1+J/vcT_k}{1-J/vcT_k} \tag{17}$$

where $T_k = E/c$ is the kinetic temperature, which in equilibrium (at $J=0$) corresponds to the thermodynamic temperature, while far from equilibrium is proportional to the local energy density. After some algebra, we obtain

$$\theta = \frac{T_k}{1 + J[\ln(1+J/cvT_k) - \ln(1-J/cvT_k)]/2cvT_k} \tag{18}$$

Parameter $\gamma$, Eq.(16), has no analog in equilibrium and must be regarded as a purely nonequilibrium quantity describing how an increment in the heat flux modifies the entropy [7,9]. In terms of the kinetic temperature $T_k$ the parameter takes the form

$$\gamma = [\ln(1 - J/vcT_k) - \ln(1 + J/cvT_k)]/2cvT_k \tag{19}$$

The derivative $\partial\theta/\partial T_k$, which will be used later to scale thermal conductivity and heat capacity to local nonequilibrium situation, is obtained from Eq.(18) as follows

$$\frac{\partial\theta}{\partial T_k} = \left(\frac{J}{cvT_k}\ln\frac{1+J/cvT_k}{1-J/cvT_k} + \frac{1}{1-(J/cvT_k)^2}\right)\left(1 + \frac{J}{2cvT_k}\ln\frac{1+J/cvT_k}{1-J/cvT_k}\right)^{-2} \tag{20}$$

*2.3 Nonequilibrium entropy and temperature as functions of the nondimensional heat flux*

Let us consider the behavior of $S_{neq}$, $\theta$, $\gamma$, and $\partial\theta/\partial T_k$ as functions of the nondimensional heat flux heat flux $q = J/J_{max}$, where $J_{max} = vcT_k$ is the maximum possible value of the heat flux, which is reached when all the heat carriers move in the same direction [6,7,9]. The physically reasonable upper bound for the heat flux $|J| \leq J_{max}$ implies that $|q| \leq 1$, i.e. $q$ takes values between $-1$ and $1$. Thus, we obtain

$$S_{neq} = S_{eq} - \frac{1}{2}(1+q)\ln(1+q) - \frac{1}{2}(1-q)\ln(1-q) \tag{21}$$

$$\theta = \frac{T_k}{1 + q[\ln(1+q) - \ln(1-q)]/2} \tag{22}$$

$$\gamma = [\ln(1-q) - \ln(1+q)]/2cvT_k \tag{23}$$

$$\frac{\partial\theta}{\partial T_k} = \left(q\ln\frac{1+q}{1-q} + \frac{1}{1-q^2}\right)\left(1 + \frac{q}{2}\ln\frac{1+q}{1-q}\right)^{-2} \tag{24}$$

The nonequilibrium entropy $S_{neq}$, Eq.(21), scaled with $S_{eq}$, is shown in Fig.1 as a function of the nondimensional heat flux $q$ (dashed line). As expected, $S$ is always less than or equal to that of at local equilibrium situation. The presence of the heat flux $|q| > 0$ reduces the value of entropy, indicating that the nonequilibrium state is more ordered than for the corresponding equilibrium state. In the maximum heat flux limit $|q| \to 1$, which corresponds to the completely ordered state, Eq.(21) gives $S_{neq} \to 0$ (see Fig.1) [6,7]. The nondimensional nonequilibrium temperature $\theta/T$, Eq.(22), as a function of the nondimensional heat flux $q$ is also shown in Fig.1 (solid line). In equilibrium $q = 0$ and, as expected, the nonequilibrium temperature $\theta$ tends to its equilibrium value $T$. The presence of the heat flux $q \neq 0$ implies that the local energy density is not completely thermalized and there is anordered fraction.

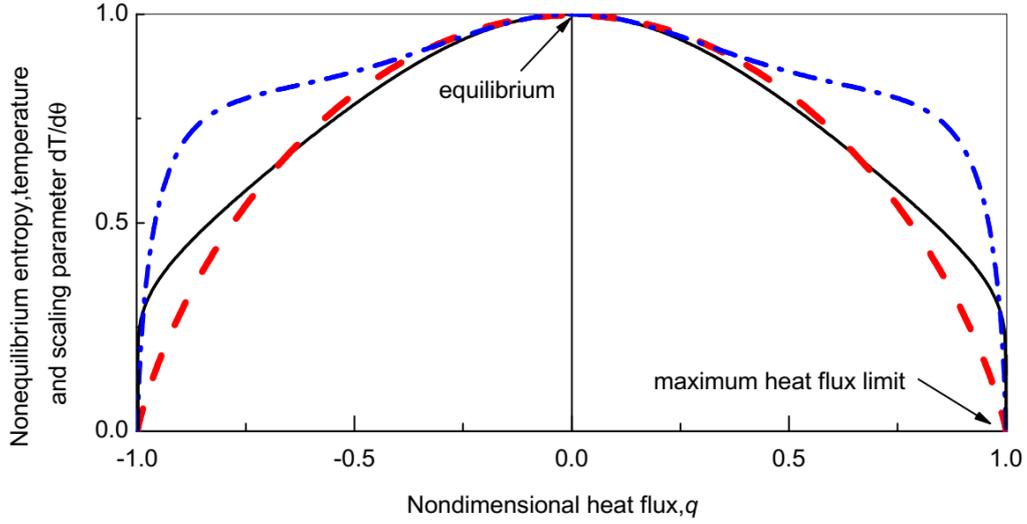

***Fig.1.*** *Nonequilibrium entropy $S_{neq}/S_{eq}$, Eq.(21), (dashed line), nonequilibrium temperature $\theta/T$, Eq.(22), (solid line) and scaling parameter $\varphi = \partial T/\partial \theta$, Eq.(24), (dash-dotted line) as functions of the nondimensional heat flux $q$.*

As the heat flux increases, this ordered fraction of the local energy density increases, whereas the disordered (thermalized) fraction decreases. Accordingly, the nonequilibrium temperature $\theta$, which is proportional to the thermalized (disordered) fraction, decreases and tends to zero in the maximum heat flux limit $q \to 1$ (see solid line in Fig.1).A similar behavior demonstrates the scaling parameter $\varphi = \partial T_k/\partial \theta$, which can be obtained from Eq.(24) (see dash-dotted line in Fig.1).

Fig.2 shows the nondimensional Lagrange multiplier $\gamma'$, Eq.(23), (dashed curve) and $1/\gamma'$ (solid curve), where $\gamma' = \gamma E$, as functions of $q$.The behavior of $\gamma$ demonstrates that the rate of change of the nonequilibrium entropy increases abruptly when $q \to 1$ (see dashed line in Fig.2).

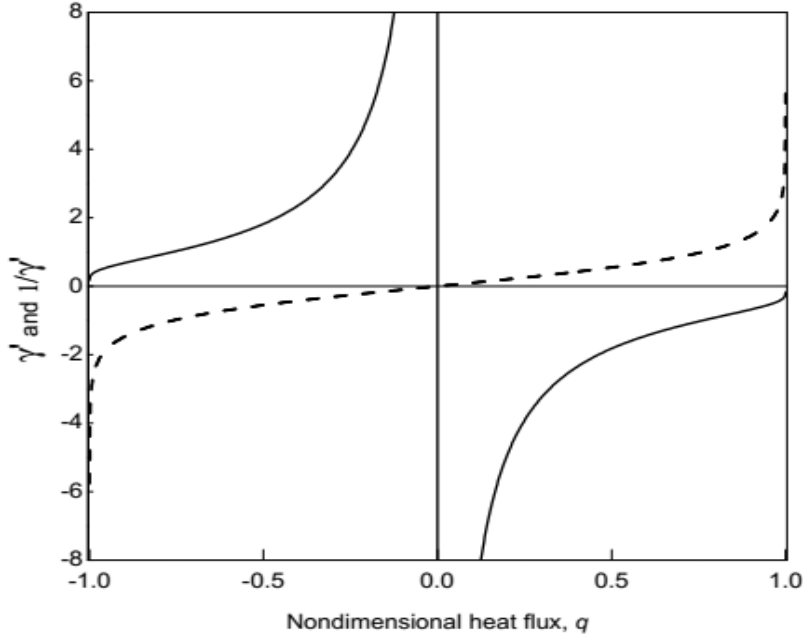

**Fig.2.** Nondimensional Lagrange multiplier $\gamma' = \gamma v E$, Eq.(23), (dashed curve) and $1/\gamma'$ (solid curve) as functions of the nondimensional heat flux $q$.

Thus, when a system equilibrates form the highly nonequilibrium state with the maximum possible value of the heat flux $q = 1$ to equilibrium state with $q = 0$, the energy of the completely ordered motion of the energy carriers converts into the thermal energy of the disordered motion. During the equilibration the ratios $S_{neq}/S_{eq}$ and $\theta/T_k$ change from zero in the completely ordered state with $q = 1$ to unity in the completely disordered, thermalized state $q = 0$. Thus, the nondimensional heat flux $q$ plays a role similar to that of an order parameter: it varies from zero in the completely disordered (thermalized) equilibrium state to unity in the completely ordered nonequilibrium state [7]. In other words, $q$ characterizes the degree of ordering, while $S_{neq}/S_{eq}$ and $\theta/T_k$ the degree of disordering or thermalization.

*2.4. Nonequilibrium entropy and temperature as functions of the kinetic temperature*

Now let us study the behavior of $S_{neq}$, $\theta$, $\gamma$, and $\partial\theta/\partial T$ as functions of the energy density $E$ (or kinetic temperature $T_k$) at constant heat flux $J$. For further consideration it is convenient to introduce the nondimensional energy density $\varepsilon = Ev/2J$ or, equivalently, the nondimensional kinetic temperature $\varepsilon = T_K cv/2J$. In this case Eqs.(21)-(24) take the form, respectively,

$$S_{neq} = S_{eq} - \frac{1}{2}\left(1 + \frac{1}{2\varepsilon}\right)\ln\left(1 + \frac{1}{2\varepsilon}\right) - \frac{1}{2}\left(1 - \frac{1}{2\varepsilon}\right)\ln\left(1 - \frac{1}{2\varepsilon}\right) \qquad (25)$$

$$\theta = \frac{\varepsilon}{1 + \left[\ln(\varepsilon + 1/2) - \ln(\varepsilon - 1/2)\right]/4\varepsilon} \qquad (26)$$

$$\gamma' = [\ln(\varepsilon - 1/2) - \ln(\varepsilon + 1/2)]/2 \tag{27}$$

$$\varphi = \left(1 + \frac{1}{4\varepsilon} \ln \frac{\varepsilon + 1/2}{\varepsilon - 1/2}\right)^2 \left(\frac{1}{2\varepsilon} \ln \frac{\varepsilon + 1/2}{\varepsilon - 1/2} + \frac{\varepsilon^2}{\varepsilon^2 - (1/2)^2}\right)^{-1} \tag{28}$$

Fig.3 shows the nonequilibrium entropy $S_{neq}$, Eq.(25), scaled with $S_{eq}$, as a function of $\varepsilon$. When $\varepsilon \gg 1$, which corresponds to the high temperature limit, deviation from equilibrium is small and as expected, gives $S_{neq}/S_{eq} \to 1$. However, as $\varepsilon$ decreases (low temperature limit), the local nonequilibrium effects begin to play an important role and decrease the entropy $S_{neq}$ such that $S_{neq} \to 0$ at $\varepsilon \to 1/2$ (see Fig.3). In dimensional units this limit corresponds to $T_K \to J/cv$.

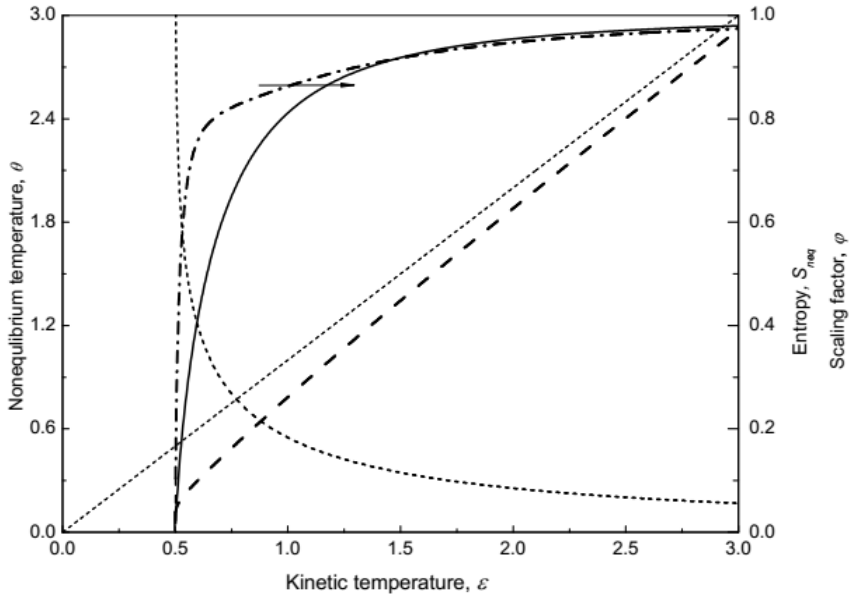

**Fig.3**. Nonequilibrium entropy $S_{neq}$, Eq.(25), scaled with $S_{eq}$, (solid line), nonequilibrium temperature $\theta$, Eq.(26), in units of $\hbar\omega/k_B$ (dashed line), nondimensional parameter $\gamma'$, Eq.(27), (short dashed line) and scaling parameter $\partial T_k/\partial \theta$ (dash-dotted line) as functions of the nondimensional kinetic temperature $\varepsilon$ for $j = 1/2$ (in units of $\hbar\omega/k_B$).

The nonequilibrium temperature $\theta$, Eq.(26), is also shown in Fig.3 as a function of $\varepsilon$. In the high temperature limit $\varepsilon \gg 1$, we obtain that $\theta \to \varepsilon$, which implies that the deviation from local equilibrium is small and LTE conditions are valid. However, in the low temperature limit, the nonequilibrium temperature $\theta$ decreases faster than the nondimensional kinetic temperature $\varepsilon$ due to increasing nonequilibrium effects (see Fig.3). When $\varepsilon \to 1/2$, Eq.(26) gives $\theta \to 0$ (see Fig.3), which corresponds to the maximum heat flux limit $q \to 1$ with completely ordered motion of energy carriers (see Fig.1).

The nonequilibrium parameter $\gamma'$, Eq.(27), has no analog in equilibrium and must be regarded as a purely nonequilibrium quantity [7,9]. As $\varepsilon$ decreases, $\gamma'$ increases (see Fig.3) describing how an increment in the deviation from equilibrium modifies the entropy.

Fig.3 shows the scaling parameter $\varphi = \partial T_k / \partial \theta$, Eq.(28), as a function of $\varepsilon$, which also implies that $\theta \to \varepsilon$ in the high temperature limit $\varepsilon \gg 1$, whereas $\theta$ differ substantially from $\varepsilon$ at low temperature $\varepsilon \to 1/2$ due to increasing nonequilibrium effects.

*2.5 Nonequilibrium thermal conductivity and heat capacity*

The heat capacity of the nonequilibrium system with heat flux is defined as $c_{neq} = (\partial E / \partial \theta)_q$, rather than $c = \partial E / \partial T_k$. The reason for this is that in a nonequilibrium state it is $\theta$ rather than $T_k$ that can be directly measured by a thermometer [9]. This definition gives

$$c_{neq} / c = (\partial T_k / \partial \theta)_q \tag{29}$$

Following [24-26,30,31,36], we assume that in the steady-state the temperature gradient is implicitly associated with the thermal conductivity such that

$$J = -\lambda_{neq} \frac{\partial \theta}{\partial x} = -\lambda \frac{\partial T_k}{\partial x} \tag{30}$$

where $\lambda$ is the thermal conductivity, which corresponds to the kinetic temperature gradient, while $\lambda_{neq}$ is the nonequilibrium thermal conductivity, which corresponds to the nonequilibrium temperature gradient. Eq.(30) allows us to rescale the thermal conductivity in addition to the heat capacity scaling, Eq.(29), as follows

$$\lambda_{neq} / \lambda = (\partial T_k / \partial \theta)_q \tag{31}$$

Thus, Eqs.(29) and (31) demonstrate that both the thermal conductivity $\lambda_{neq} / \lambda$ and the heat capacity $c_{neq} / c$ have the same scaling due to nonequilibrium effects $\varphi = (\partial T / \partial \theta)_q$, Eq.(20). The scaling implies that in the low temperature limit $T_k \to j$ both $\lambda_{neq} \to 0$ and $c_{neq} \to 0$, while in the high temperature limit $T_k \gg j$, $\lambda_{neq} \to \lambda$ and $c_{neq} \to c$ (see Fig.3).

**3. Comparison with quantum systems**

*3.1. Quantum systems with bounded energy spectrum*

The nonequilibrium entropy $S_{neq}$, Eq.(21), as a function of the nondimensional heat flux $q$ (see Fig.1) coincides with the equilibrium entropy of spin system in magnetic field as a function of the nondimensional energy $U/N\sigma$, where $U$ is the energy of the spin system, $N$ is the number of particles, $\sigma$ is the energy difference between the quantum states of the spins [38]. Both entropies increase from zero at $q = U/N\sigma = -1$ up to their maximum values $S_{max} = S_{eq}$ at $q = U/N\sigma = 0$ and then decreases to zero at $q = U/N\sigma = 1$ (compare Fig.1 with Fig.3.14 in Ref.[38]). The analogy arises because both systems consist of two interaction subsystems, namely, the 1D nonequilibrium system with heat flux consists of the two groups of energy carries moving in the opposite directions, while the equilibrium spin system consists of the two groups with spins $1/2$ and $-1/2$. The heat flux in the present model $J \propto (E_1 - E_2)$, where $E_1$ and $E_2$ are the energy densities of the carriers moving to the right and to the left, respectively, while the energy of the spin system $U \propto (U_1 - U_2)$, where $U_1$ and $U_2$ are the energy of subsystems with different spins. This implies that the nonequilibrium entropy, Eq.(21), depends on the nondimensional heat flux $q$ in a similar way as the equilibrium entropy of the spin system depends on the nondimensional energy $U/N\sigma$ [38,39].

The effective spin temperature $T_S$ is defined as $T_S = (\partial S/\partial U)^{-1}$ [38,39]. The corresponding derivative in the present model $(\partial S/\partial q)^{-1} = \gamma^{-1}$ can be obtained from Eq.(23) and is shown in Fig.2 as a function of $q$. The inverse of $\gamma$ exhibits a singularity and changes sign as $q$ crosses $q = 0$, which corresponds to the behavior of $T_S$ as a function of $U$ in systems with bounded energy spectrum (compare, for example, Fig.2 and Fig.3.15 in Ref.[38]).

Thus, some analogies between the behavior of the nonequilibrium system with heat flux and the equilibrium spin system in magnetic field arise due to the bounded spectrum of the heat flux $-1 \leq q \leq 1$.

*3.2 Quantum harmonic oscillator*

Mean energy of one quantum harmonic oscillator is given as [38,40,41]

$$E = \frac{1}{2}\hbar\omega + \frac{\hbar\omega}{\exp(\hbar\omega/k_B T) - 1} \qquad (32)$$

where $T$ is temperature of the oscillator, $k_B$ is Boltzmann constant, $\omega$ is frequency of the oscillator, $\hbar$ is Planck constant divided by $2\pi$. At high temperatures, the energy of the harmonic oscillator is $kT$, which is the classical result, as expected. Meanwhile, at low temperatures, the energy is asymptotically $E_{ZP} = \hbar\omega/2$ (the first term on the right hand side of Eq.(32)), which is the so-called "zero-point" energy (or zero-point temperature) associated with quantum fluctuations. For further consideration it is more convenient to represent the temperature of the quantum harmonic oscillator $T$ as a function of the mean energy $E$, which is given by

$$T = \frac{\hbar\omega}{k_B}\left(\ln\frac{E+\hbar\omega/2}{E-\hbar\omega/2}\right)^{-1} \qquad (33)$$

Since the zero-point energy is associated with quantum fluctuation, the corresponding value of the heat flux in the present model is connected to the zero-point energy as follows $J_{ZP} = vE_{ZP} = v\hbar\omega/2$, which is the smallest value of heat flux where the thermodynamic temperature $T$ falls to zero.

Note that the nonequilibrium temperature $\theta$ in the present model plays a similar role as the thermodynamic temperature of the quantum harmonic oscillator $T$, while the kinetic temperature $T_k$ characterizes local energy density and corresponds to the mean energy of the quantum harmonic oscillator $E$. This allows us to compare the behavior of the temperatures as functions of the local energy, which is shown in Fig.4.

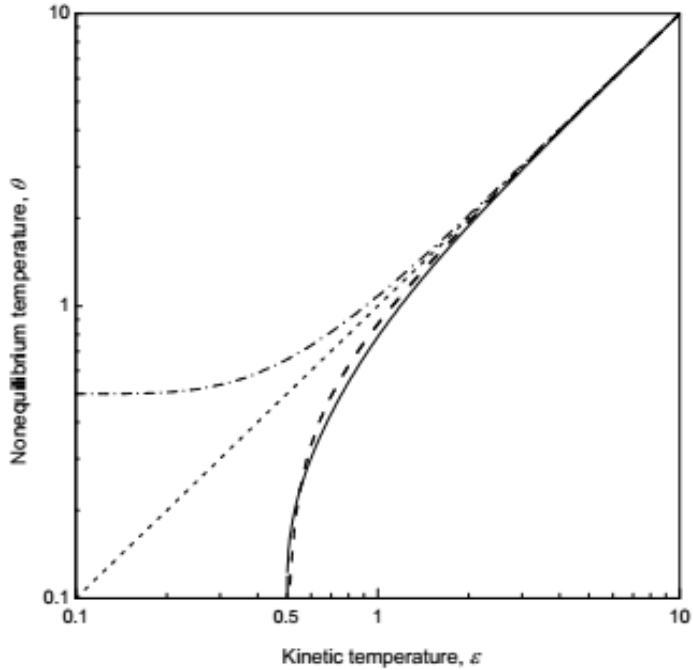

**Fig.4.** Nonequilibrium temperature $\theta$, Eq.(34), for $j = 1/2$ (solid line) and quantum oscillator temperature, Eq.(32), (dashed line) in units of $\hbar\omega/k_B$ as functions of the nondimensional energy $\varepsilon = E/\hbar\omega$. Dash-dotted line shows $E$ as a function of $\theta$.

We observe that the nonequilibrium temperature $\theta$, Eq.(26), and the quantum harmonic oscillator temperature $T$, Eq.(33), demonstrate similar behavior as functions of $\varepsilon = vE/2J_{ZP} = E/\hbar\omega$. In the high energy limit $\varepsilon \to \infty$, the temperatures coincide with the energy density $\varepsilon$, which is the classical result, as expected. Meanwhile, as $\varepsilon$ decreases, the temperatures go to zero at the nonzero value of the energy $\varepsilon_{ZP} = 1/2$. Thus, the heat flux in the nonequilibrium system (expressed in corresponding units) plays a role of the zero-point energy - it bounds the energy spectrum of the nonequilibrium system when $\theta \to 0$.

The heat capacity of the quantum harmonic oscillator as a function of $\varepsilon$ is given by [38,40,41]

$$c_{osc}/k_B = (\varepsilon^2 - 1/4)[\ln(\varepsilon + 1/2) - \ln(\varepsilon - 1/2)]^2 \qquad (34)$$

Comparison of the heat capacities $c_{osc}$, Eq.(37), and $c_{neq}$, Eq.(28), as functions of $\varepsilon$ is shown in Fig.5.

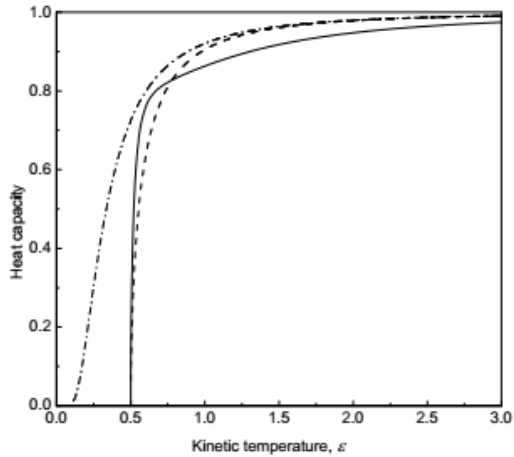

**Fig.5.** Nonequilibrium heat capacity $c_{neq}$, Eq.(28), (solid line) and quantum oscillator the heat capacities $c_{osc}$, Eq.(37), (dashed line) as functions of $\varepsilon = E/\hbar\omega$. Dashed dotted line represents $c_{osc}$ as function of the oscillator temperature in units of $\hbar\omega$.

Here again, we obtain some similarities in behavior of the effective nonequilibrium heat capacity $c_{neq}$ and the quantum heat capacity $c_{osc}$, which both tend to zero at $\varepsilon \to 1/2$ and go to their equilibrium high temperature values at $\varepsilon \to \infty$.

It should be mentioned that Nagata [41] obtained equilibrium thermodynamic quantities, such as temperature, entropy, and heat capacity, in quantum harmonic oscillator model as functions of the thermal average of the uncertainty relation $(\Delta x)_T (\Delta p_x)_T$. This average is carried out on the basis of the fluctuation analysis in canonical system in thermal equilibrium with a heat reservoir at temperature $T$. The equilibrium thermodynamic quantities of Nagata as functions of the thermal uncertainty relation behave similar to their nonequilibrium counterparts in the present model as functions of $\varepsilon$ (compare, for example, Fig.5 and Fig.6 in Ref.[41] with Fig.5 and Fig.6 in the present paper, respectively). Nagata stated that the thermal uncertainty relation in his model seems to play a role just like temperature, over the region of $(\Delta x)_T (\Delta p_x)_T > 1$. As discussed above, the nonequilibrium thermodynamic quantities can be described as functions of $\varepsilon$ in the region $\varepsilon \geq 1/2$. This implies that the thermal uncertainty relation of Nagata can be treated as the kinetic temperature, which reflects the mean energy in the region $(\Delta x)_T (\Delta p_x)_T \geq 1/2$.

*3.3 Quantum correction in MD simulations*

*3.3.1 Temperature rescaling*

MD simulations are widely and successfully used to study heat conduction in nano level [1,2,23-37]. However, the MD approach is not suitable at low temperature when quantum effects on the phonon mode populations are important. In a classical system at a given temperature, all modes are excited approximately equally, whereas in the quantum system, there is a freezing out of high-frequency modes at low temperatures [24,38]. The main idea to overcome this difficulty and include quantum effects in the MD simulations is to scale the MD temperature $T_{MD}$ and introduce the ''real'' quantum temperature $T$ using simple quantum-mechanical calculations and/or arguments. The temperature of the "real" quantum system can be found by equating the kinetic energy in MD simulations to that of a quantum phonon system, such that [23-37]

$$3(N-1)k_B T_{MD} = \sum_k \left( \frac{\hbar \omega_k}{2} + \frac{\hbar \omega_k}{\exp(\hbar \omega_k / k_B T) - 1} \right) \qquad (35)$$

where summation is over the $k$ normal modes of the system. Essentially, this procedure provides a means for mapping results calculated classically onto their quantum analogs at the same energy level. The idea behind this rescaling scheme is that one hopes to establish a one-to-one correspondence between the real quantum system and the classical MD simulation, such that all physical observables are the same. The calculations [23-37] demonstrate that the $T_{MD}$ curve approaches $T$ as the temperature is increased and more modes are excited, which implies the classical limit. However, as the temperature is decreased, the difference between the temperatures becomes significant due to the quantum effects, namely, the quantum corrected temperature $T$ decrease faster in comparison with $T_{MD}$ and reaches zero at $T_{MD} > 0$ due to the "zero point temperature" [23-38]. The behavior of $T$ as a function of $T_{MD}$ is similar to the behavior of the temperature of the quantum harmonic oscillator $T$ as a function of its energy $E$, discussed above. This is not surprising because $T$ is calculated from Eq.(35), which represents a sum of the energies of quantum harmonic oscillators with different frequencies. Thus, the behavior of the nonequilibrium temperature $\theta$ as a function of the kinetic temperature $T_k$ in the present model is similar not only to the behavior of the temperature of the quantum harmonic oscillator, as discussed above, but also to the behavior of the real quantum temperature $T$ as a function of the kinetic temperature $T_{MD}$ in the MD simulations of different systems [23-37]. The only difference is a value of the zero point energy, which bounds the energy spectrum of the quantum system.

Fig.6 shows the MD calculations of Gomes et al. [36] and Wang et al. [28] of the quantum corrected temperature $T$ of silicon, which are in good agreement with the nonequilibrium temperature $\theta$, Eq.(18). Thus a significant leveling off in the maximum heat flux limit $T \to |j|$ behaves similar to the real temperature of equilibrium system in the low temperature quantum limit $T$, such that the normalized heat flux expressed in degrees kelvin $j$ plays a role analogues to the zero point temperature.

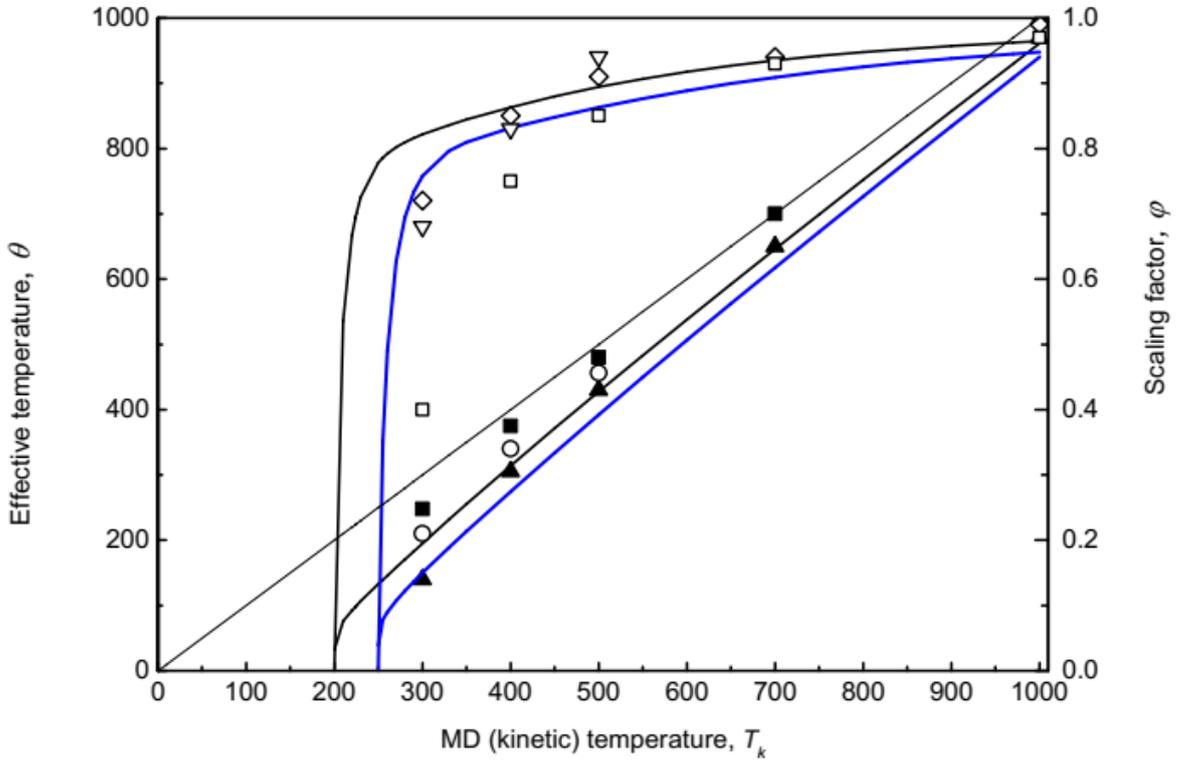

**Fig.6.** Effective temperature $\theta$, Eq.(26), (solid line) and the scaling factor $\partial T / \partial \theta$, Eq.(28), (dashed line) as functions of $T$ for $j = 200\,\text{K}$ (black lines) and $j = 250\,\text{K}$ (blue lines). Points are the MD calculations of Gomes et al.[36] ($T_{real}$ - solid triangles and squares, $\partial T_{MD}/\partial T_{real}$ - open squares and diamonds) and Wang et al.[28] ($T_{real}$ - open circles, $\partial T_{MD}/\partial T_{real}$ - open triangles).

Fig.7 shows the MD calculation of the quantum temperature $T$ for perfect SiC crystal in comparison with the nonequilibrium temperature $\theta$, Eq.(18), which also demonstrates a good agreement between the quantum corrected temperature of equilibrium system and the extended nonequilibrium temperature in system with heat flux.

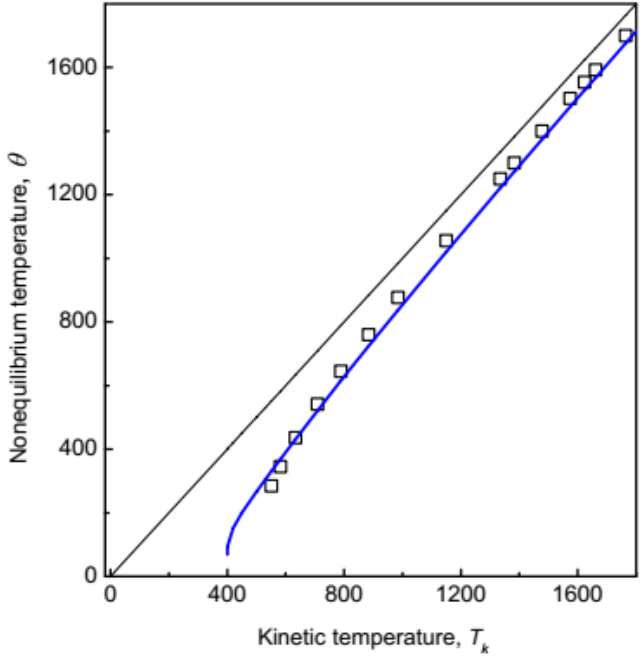

**Fig.7.** Effective temperature $\theta$, Eq.(26), (solid line) as a functions of the kinetic temperature $T_k$ for $j = 400\,\text{K}$. Points are the MD calculations of Li et al. [35] for perfect SiC crystals.

*3.3.2. Thermal conductivity rescaling*

In addition to the temperature rescaling in MD simulations, the temperature gradient, $\partial T_{MD}/\partial T$, multiplies the thermal conductivity since the gradient is implicitly associated with the thermal conductivity (see Eq.(24)), which rescales the thermal conductivity as follows [24-26,28-31,36].

$$\frac{\lambda_{real}}{\lambda_{MD}} = \frac{\partial T_{MD}}{\partial T} \tag{36}$$

The thermal conductivity scaling factor $\partial T_{MD}/\partial T$, Eq.(36), has been considered as a function of $T$ in Refs.[24,25,30] and as a function of $T_{MD}$ in Refs.[26,29,30]. As the temperatures increase, the scaling factor, Eq.(28), approaches unity in both cases, which demonstrates negligible quantum effects in the classical high temperature limit. At low temperature the scaling factor, Eq.(36), decreases and reaches zero as a function of $T$ when $T = 0$ [5,20]. As a function of $T_{MD}$ the scaling factor, Eq.(36), reaches zero at $T_{MD} > 0$ due to the "zero point temperature" [26,30]. Thus, the quantum correction $\partial T_{MD}/\partial T$ in the MD simulations and the scaling factor in the present model $\varphi = \partial T/\partial \theta$ exhibit a qualitatively similar behavior (compare, for example, Fig.3 with Fig.1.3 in Ref.[30], Fig.2 in Ref.[29], and Fig.3 in Ref.[26]). Moreover, the similarity is quantitatively confirmed by Fig.3, which shows that the MD calculations of $\partial T_{MD}/\partial T$ by Gomes et al.[36] and Wang et al.[28] are in good agreement with the scaling factor $\varphi = \partial T / \partial \theta$, Eq.(28).

*3.3.3. Heat capacity rescaling*

The classical nature of the MD simulations is perhaps most evident when considering the predicted heat capacity, and how they differ from the quantum-mechanical calculations. As expected, the classical and quantum harmonic specific heats are significantly different at low temperatures, where quantum effects are important [24,32,34]. The classical theoretical prediction of the heat capacity does not depend on temperature, whereas the quantum calculation demonstrates that the specific heat decreases with decreasing temperature [5,37,40]. To take into account the quantum effects at low temperatures, the heat capacity is defined as $c = \partial E / \partial T$, rather than $c = \partial E / \partial T_{MD}$ [34,35], which implies that the scaling factor $\partial T_{MD} / \partial T$ plays a similar role in the calculations of the heat capacity as in the calculations of the thermal conductivity (see previous section).

Note that the definition of heat capacity with allowance for quantum effects $c = \partial E / \partial T$ corresponds to the definition of heat capacity for nonequilibrium systems $c_{neq} = \partial E / \partial \theta$ because just the effective temperature $\theta$ of nonequilibrium systems plays a role similar to the real temperature $T_{real}$ of equilibrium systems in the low temperature quantum limit.

*3.3.4. Comments*

The study developed in the previous sections demonstrates some analogies between the behavior of the nonequilibrium system in the maximum heat flux limit and the equilibrium quantum system at low temperatures. The mean energy per quantum oscillator is different from $k_B T$, actually, it is always greater than $k_B T$ (see Eqs.(32)-(33) and Fig.4). This implies that the quantum-mechanical oscillators do not obey the equipartition theorem [38]. Only in the limit of high temperatures, where the thermal energy $k_B T$ is much larger than the energy quantum $\hbar \omega$, the mean energy per oscillator *E* tends to the equipartition value. The equipartition theorem is not always valid; it applies only when the relevant degrees of freedom can be *freely* excited [38]. However, at a given temperature *T*, there may be certain degrees of freedom which, due to the insufficiency of the energy available, are more or less "frozen" due to quantum mechanical effects. Thus, the difference between $k_B T$ and $E$ (or $k_B T_{MD}$) in the quantum systems arises due to the freezing out of high-frequency modes at low temperatures [24,30,38]. The freezing out violates the equipartition and decreases the real quantum temperature $T$ in comparison with its equipartition value $E/k_B$ (or $T_{MD}$) such that the difference increases at low temperature due to increasing quantum effects [23-38]. A similar situation occurs in the nonequilibrium systemswhere the breakdown of equipartition implies that the generalized nonequilibrium temperature $\theta$, Eq.(18), is always less or equal to the kinetic temperature $T_k$. The nonequilibrium temperature $\theta$, defined as the inverse of the derivative of the nonequilibrium entropy with respect to the energy, is proportional to the thermalized fraction of the local energy density, whereas the kinetic temperature $T_k$ is proportional to the local energy density, which is the sum of the thermalized and non-thermalized fractions. In other words, the non-thermalized (ordered) fraction, that guarantees the nonzero heat flux, does not contribute to the nonequilibrium temperature $\theta$. This looks like a "freezing out" of the non-thermalized fraction of the local energy density when the deviation from equilibrium is high, similar to the freezing out of high-frequency modes in the equilibrium quantum systems at low temperature. Thus, the analogies between the behavior of nonequilibrium systems in the maximum heat flux limit and equilibrium quantum systems at low temperatures arise due to the breakdown of equipartition when some degrees of freedom do not make a significant contribution toward the thermalized (internal) energy of the system.

**Conclusion**

Using information entropy approach, we extend the meaning of equilibrium variables, such as entropy, temperature, heat capacity, and thermal conductivity to the nonequilibriumscenario when 1D heat conduction is described by the Boltzmann transport equation with single relaxation time approximation. In the maximum heat flux limit $J \to J_{max}$, the nonequilibrium entropy, temperature, heat capacity, and thermal conductivity tend to zero even at nonzero kinetic temperature $T_k$, which is proportional to the local energy density. This provides possible generalization of the third law to the nonequilibrium case. The nondimensional heat flux $q = J/J_{max}$ plays a role of an order parameter: it changes from zero in the disordered equilibrium state $J = 0$ to unity in the completely ordered nonequilibrium state $J = J_{max}$.

Some analogies between the behavior of nonequilibrium systems in the maximum heat flux limit and equilibrium quantum systems at low temperatureshave been observed. In both cases the local energy density expressed in degrees kelvin exceeds the (nonequilibrium) temperature due to the breakdown of equipartition.The normalized heat flux expressed in energy units plays a role similar to the zero-pointenergy in quantum systems. The observed analogies imply that there can be fruitful cross-fertilization of ideas and techniques between these two fields.

The purpose of this paper is not so much to provide exact expressions for the thermodynamic quantities but to emphasize their most prominent qualitative features as one tends to the maximum deviation from equilibrium. From this perspective, the present work provides very compact expressions on a much simpler basis than existing theories and, hence, it is worthwhile to use them for a conceptual exploration or as an effective tool for rapid calculations to make more elaborated approaches, such as MD simulations or BTE, less computationally expensive.

## ACKNOWLEDGMENTS

The study was supported by the Ministry of Education and Science of the Russian Federation in the framework of the State Task (project no. 0089-2019-0002).